\documentclass[twocolumn,aps,prd,10pt,superscriptaddress,longbibliography]{revtex4-1}
\usepackage{amsmath}
\usepackage{bm}
\usepackage{graphicx}
\usepackage{multirow}
\usepackage[usenames,dvipsnames,svgnames]{xcolor}
\usepackage{hyperref}
\hypersetup{
pdfnewwindow=true,      
colorlinks=true,        
linkcolor=Blue,         
citecolor=Blue,         
filecolor=Blue,         
urlcolor=Blue           
}

\begin{document}

\title{Improving the accuracy of the neuroevolution machine learning potential for multi-component systems}

\author{Zheyong Fan}
\email{brucenju@gmail.com}
\affiliation{College of Physical Science and Technology, Bohai University, Jinzhou 121013, P. R. China}
\date{\today}

\begin{abstract}
In a previous paper [Fan Z \textit{et al}. 2021 Phys. Rev. B, \textbf{104}, 104309], we developed the neuroevolution potential (NEP), a framework of training neural network based machine-learning potentials using a natural evolution strategy and performing molecular dynamics (MD) simulations using the trained potentials. The atom-environment descriptor in NEP was constructed based on a set of radial and angular functions. For multi-component systems, all the radial functions between two atoms are multiplied by some fixed factors that depend on the types of the two atoms only. In this paper, we introduce an improved descriptor for multi-component systems, in which different radial functions are multiplied by different factors that are also optimized during the training process, and show that it can significantly improve the regression accuracy without increasing the computational cost in MD simulations. 
\end{abstract}

\maketitle

\section{Introduction}

In recent years, machine-learning (ML) potentials \cite{behler2016jcp,Deringer2019am,Mueller2020jcp,Mishin2021am,Unke2021cr} have played an important role in molecular dynamics (MD) simulations. A well trained ML potential can achieve an accuracy close to that of the training data and a speed that cannot be achieved by \textit{ab initio} MD simulations. After the pioneering work by Behler and Parrinello \cite{behler2007prl} on the high-dimensional neural network (NN) potential, other alternatives such as the Gaussian approximation potential (GAP) \cite{bartok2010prl} and some linear regression ML potentials \cite{Thompson2014jcp,Shapeev2016} were also developed. Many methods and computer codes for constructing NN potentials  have been developed by exploring standard ML libraries \cite{wang2018cpc,zhang2018prl,lee2019cpc,lot2020cpc,gao2020jcim,shao2020jcim,Pattnaik2020jpca,Yanxon2021,zhang2021prl}. 

Recently, the present author developed a framework called neuroevolution potential (NEP) \cite{fan2021neuroevolution} for training NN-based ML potential using a natural evolution strategy \cite{Schaul2011,wierstra2014jmlr}, instead of the conventional back propagation (gradient descent) approach. NEP has been implemented in version $2.6$ of the open-source \textsc{gpumd} package \cite{fan2013cpc,fan2017cpc,gpumd-github}. It has been demonstrated \cite{fan2021neuroevolution} that NEP as implemented in \textsc{gpumd} can achieve an accuracy comparable to other popular implementations of ML potentials \cite{quip,Novikov2021,wang2018cpc}, while exhibiting a much higher computational efficiency in MD simulations.

In this paper, we show that for multi-component systems, i.e., systems with multiple atom types, the accuracy of NEP can be significantly improved. We present the improved approach and implement it version $2.9$ of \textsc{gpumd}. For simplicity, the NEPs as implemented in versions $2.6$ and $2.9$ of \textsc{gpumd} will be called NEP1 and NEP2, respectively. We will use bulk PbTe and Al-Cu-Mg alloy as case studies to show the improved accuracy of NEP2 as compared to NEP1. 

\section{Theory}

\subsection{The previous NEP1}

The ML potential in NEP1 \cite{fan2021neuroevolution} is a local many-body one, where ``local'' means that the total potential energy $U$ of a system of $N$ atoms can be written as a sum of site energies, $U = \sum_{i=1}^N U_i$. The site energy $U_i$ of atom $i$ is taken as a function of a set of $N_{\rm des}$ descriptor components $\{q^i_{\nu}\}_{\nu=1}^{N_{\rm des}}$. The function is taken as a NN with a single hidden layer with $N_{\rm neu}$ neurons: 
\begin{equation}
\label{equation:nn2}
U_i = \sum_{\mu=1}^{N_{\rm neu}}w^{(2)}_{\mu}\tanh\left(\sum_{\nu=1}^{N_{\rm des}} w^{(1)}_{\mu\nu} q^i_{\nu} - b^{(1)}_{\mu}\right) - b^{(2)},
\end{equation}
where $w^{(1)}_{\mu\nu}$, $w^{(2)}_{\mu}$, $b^{(1)}_{\mu}$, and  $b^{(2)}$ are the trainable weight and bias parameters in the NN.

For a central atom $i$, there is a set of radial descriptor components ($0\leq n\leq n_{\rm max}^{\rm R}$),
\begin{equation}
\label{equation:qin}
q^i_{n}
= \sum_{j\neq i} g_n(r_{ij}),
\end{equation}
and a set of angular descriptor components ($0\leq n\leq n_{\rm max}^{\rm A}$ and $1\leq l\leq l_{\rm max}$),
\begin{equation}
q^i_{nl} 
= \sum_{j\neq i}\sum_{k\neq i} g_n(r_{ij}) g_n(r_{ik})
P_l(\cos\theta_{ijk}),
\label{equation:qinl}
\end{equation}
where $P_l(\cos\theta_{ijk})$ is the Legendre polynomial of order $l$, $\theta_{ijk}$ being the angle formed by the $ij$ and $ik$ bonds. The functions $g_n(r_{ij})$ are radial functions and they are defined as
\begin{equation}
\label{equation:g_n}
g_n(r_{ij}) = \frac{T_n\left(2\left(\frac{r_{ij}}{r_{\rm c}}-1\right)^2-1\right)+1}{2} f_{\rm c}(r_{ij}) c_{ij}.
\end{equation}
Here $T_n(x)$ is the $n$-th order Chebyshev polynomial of the first kind and $f_{\rm c}(r_{ij})$ is the cutoff function defined as
\begin{equation}
   f_{\rm c}(r_{ij}) 
   = \frac{1}{2}\left(
   1 + \cos\left( \pi \frac{r_{ij}}{r_{\rm c}} \right) 
   \right) 
\end{equation}
for $r\leq r_{\rm c}$ and $f_{\rm c}(r_{ij}) =0$ for $r > r_{\rm c}$. The cutoff radius $r_{\rm c}$ can take different values for the radial and angular components, which are denoted as $r_{\rm c}^{\rm R}$ and $r_{\rm c}^{\rm A}$, respectively.

Following Refs. \onlinecite{Gastegger2018jcp,Artrith2017prb}, a factor $c_{ij}$ is included in the definition of the radial functions $g_n(r_{ij})$ to account for the different atom types. Gastegger \textit{et al.} \cite{Gastegger2018jcp} suggested to use $c_{ij}=z_j$, where $z_j$ is the atomic number of atom $j$ and Artrith \textit{et al.} \cite{Artrith2017prb} suggested to use $c_{ij}=\pm 1, \pm 2, \cdots$. In NEP1, $c_{ij}$ is chosen as $\sqrt{z_iz_j}$. 

\subsection{The improved NEP2}

It is clear that in all the schemes above, the resulting descriptor has the permutation symmetry, i.e., the descriptor is invariant upon a permutation of the atoms with the same type. However, hand-chosen values for $c_{ij}$ might not be optimal. More importantly, the coefficients $c_{ij}$ are the same for all the radial functions $g_n(r_{ij})$, which do not depend on $n$. In NEP2, we propose to make these coefficients $n$-dependent, leading to the following radial functions:
\begin{equation}
\label{equation:g_n_nep2}
g_n(r_{ij}) = \frac{T_n\left(2\left(\frac{r_{ij}}{r_{\rm c}}-1\right)^2-1\right)+1}{2} f_{\rm c}(r_{ij}) c_{nij}.
\end{equation}
If the considered material has $N_{\rm typ}$ atom types, the number of $c_{nij}$ coefficients is
\begin{equation}
N_{\rm typ}^2 
\left(
n_{\rm max}^{\rm R} + n_{\rm max}^{\rm A} + 2
\right).
\end{equation}
The factor $N_{\rm typ}^2$ enumerates all the possible ordered combinations of atom types: both atoms $i$ and $j$ can be one of the $N_{\rm typ}$ types. Therefore, for a given $n$, there are $N_{\rm typ}^2$ possible $c_{nij}$ values. Taking $N_{\rm typ}=3$ as an explicit example and denoting the atom types as $a$, $b$, and $c$, we have the following $N_{\rm typ}^2=9$ coefficients for a given $n$: $c_{naa}$, $c_{nab}$, $c_{nac}$, $c_{nba}$, $c_{nbb}$, $c_{nbc}$, $c_{nca}$, $c_{ncb}$, and $c_{ncc}$; there will be $9 
\left(
n_{\rm max}^{\rm R} + n_{\rm max}^{\rm A} + 2
\right)$ parameters in total. These $c_{nij}$ parameters are not hand chosen, but are taken as free parameters to be optimized during the training process, similar to the weight and bias parameters in the NN. One difference between the $c_{nij}$ parameters and the NN parameters is that we require that
\begin{equation}
|c_{nij}| \geq \frac{1}{10}.
\end{equation}
The purpose of applying this restriction is to avoid too small values for the descriptor components. 

The method can also be applied to neural network potentials trained using the back propagation method, but one needs to compute the derivatives of a loss function with respect to the extra parameters introduced into the descriptor. By contrast, the introduction of these parameters adds little extra work in our approach as the natural evolution strategy does not require the calculation of the derivative of the loss function with respect to any parameter. This is one of the advantages of the natural evolution strategy compared to back propagation. In the moment tensor potential (MTP) \cite{Gubaev2019cms}, similar optimization of some radial coefficients has been used for multi-component systems. In the recursively embedded-atom NN potential \cite{zhang2021prl}, these coefficients for a central atom $i$ are considered to be recursively dependent on the descriptors of the neighbor atoms $j$ through extra NNs. 

We stress that the introduction of more parameters to the radial functions does not add more computations to the ML potential, as the number of descriptor components and the number of NN parameters, which affect the speed of the potential in MD simulations, are not changed. With this in mind, we next evaluate the regression accuracy of NEP2 as compared to NEP1.  

\section{Results and discussion \label{section:performance}}

\subsection{Bulk PbTe}

We first use the training data set of bulk PbTe as studied in Ref. \onlinecite{fan2021neuroevolution} to compare NEP1 and NEP2. There are 325 structures, each with 250 atoms. The training data set and related inputs and outputs are available from a public Gitlab repository \cite{nep-data}.
All the relevant hyperparameters are the same for NEP1 and NEP2 and are listed in Table \ref{table:nep-hyper}.

\begin{table}[thb]
\centering
\setlength{\tabcolsep}{2Mm}
\caption{The hyperparameters used in the NEP potential for the two materials, bulk PbTe and Al-Cu-Mg alloy. Here, $r_{\rm c}^{\rm R}$ ($r_{\rm c}^{\rm A}$) is the cutoff radius for the radial (angular) components of the descriptor, $n_{\rm max}^{\rm R}$ ($n_{\rm max}^{\rm A}$) is the Chebyshev polynomial expansion order for the radial (angular) components, $l_{\rm max}$ is the Legendre polynomial expansion order for the angular components, $N_{\rm neu}$ is the number of neurons in the hidden layer of the neural network, $\lambda_1$ ($\lambda_2$) is the $\mathcal{L}_1$ ($\mathcal{L}_2$) regularization parameter, $N_{\rm bat}$ is the batch size (number of structures used within one generation). $N_{\rm pop}$ is the population size in the natural evolution strategy algorithm, and $N_{\rm gen}$ is the maximum number of generations to be evolved. The training time using one GeForce RTX 2080ti GPU is also provided.}
\label{table:nep-hyper}
\begin{tabular}{llll}
\hline
\hline
Parameter & PbTe & Al-Cu-Mg alloy    \\
\hline
$r_{\rm c}^{\rm R}$  & 8 \AA & 6 \AA   \\
$r_{\rm c}^{\rm A}$  & 4 \AA & 4 \AA  \\
$n_{\rm max}^{\rm R}$  & $12$ & $15$   \\
$n_{\rm max}^{\rm A}$  & $6$ & $10$  \\
$l_{\rm max}$  & $4$  & $4$  \\
$N_{\rm neu}$  & $40$ & $40$  \\
$\lambda_{1}$  & $0.05$ & $0.05$  \\
$\lambda_{2}$  & $0.05$ & $0.05$  \\
$N_{\rm bat}$  & $25$ & $1000$  \\
$N_{\rm pop}$  & $50$ & $50$  \\
$N_{\rm gen}$  & $10^5$ & $10^5$ \\
Training time  & 1 hour & 7 hours \\
\hline
\hline
\end{tabular}
\end{table}

\begin{figure}[htb]
\begin{center}
\includegraphics[width=\columnwidth]{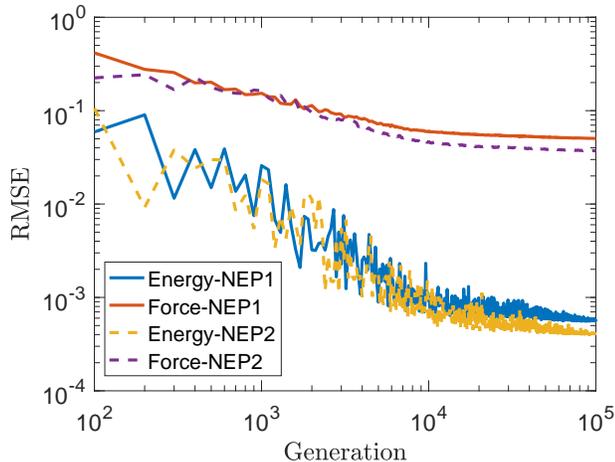}
\caption{Evolution of the energy and force RMSEs for bulk PbTe during the training process for NEP1 (solid lines) and NEP2 (dashed lines). }
\label{figure:loss}
\end{center}
\end{figure}

Figure \ref{figure:loss} shows the evolution of the root mean square errors (RMSEs) of energy and force as obtained by NEP1 and NEP2 with respect to the generation in the natural evolution strategy. The total number of generations is chosen as $10^5$ here, which is large enough to achieve convergence of the RMSEs. Within the first few thousand generations, the RMSEs are comparable between NEP1 and NEP2. However, NEP2 develops smaller RMSEs afterwards. Up to $10^5$ generations, the energy and force RMSEs obtained in NEP1 are $0.56$ meV/atom and $50$ meV/\AA~ respectively. The corresponding values obtained in NEP2 are $0.39$ meV/atom and $38$ meV/\AA. The reduction of regression errors is about $30\%$ for both energy and force. For completeness, we list the RMSEs and mean absolute errors (MAEs) of all the relevant potentials in Table \ref{table:error}.

\begin{table}[thb]
\centering
\setlength{\tabcolsep}{2Mm}
\caption{Accuracy comparison between NEP1 and NEP2. Energy and virial errors are in units of meV/atom, and force error is in units of meV/\AA. }
\label{table:error}
\begin{tabular}{lllll}
\hline
\hline
Material & Accuracy  & NEP1 & NEP2 \\
\hline
\multirow{4} {*} {Bulk PbTe}  
& Energy RMSE   & $0.56$  & $0.39$  \\
& Energy MAE    & $0.37$  & $0.29$  \\
& Force RMSE    & $50$    & $38$  \\
& Force MAE     & $36$    & $28$   \\
\hline
\multirow{6} {*} {Al-Cu-Mg alloy} 
& Energy RMSE   & $510$  & $11$  \\
& Energy MAE    & $410$  & $7.3$  \\
& Force RMSE    & $400$  & $72$  \\
& Force MAE     & $260$  & $50$   \\
& Virial RMSE   & $330$  & $43$  \\
& Virial MAE    & $190$  & $27$   \\
\hline
\hline
\end{tabular}
\end{table}

\begin{figure}[htb]
\begin{center}
\includegraphics[width=\columnwidth]{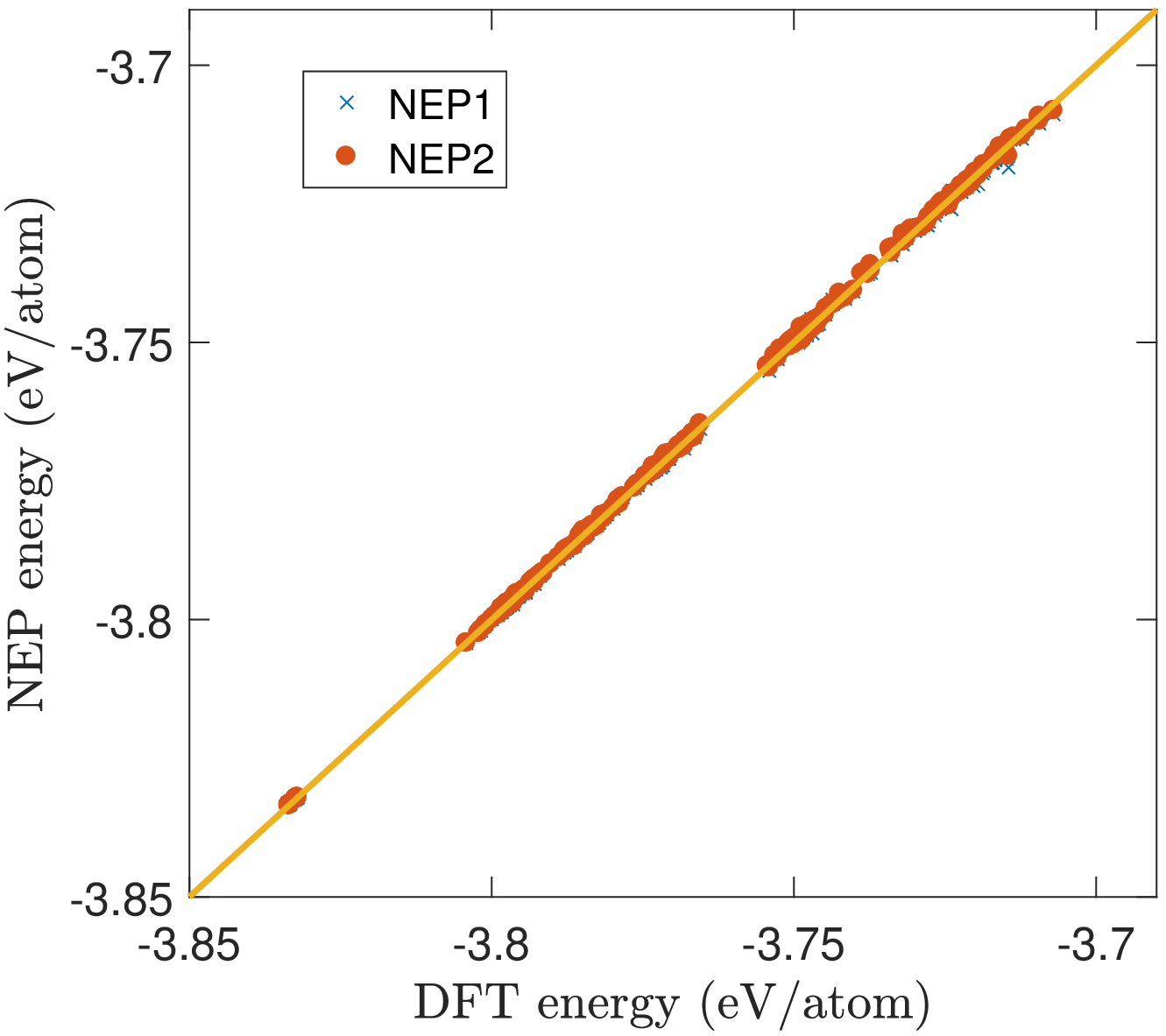}
\caption{Energy as calculated from NEP1 and NEP2 for bulk PbTe compared with the training data from quantum mechanical DFT calculations. The solid line represents the identity function used to guide the eyes.}
\label{figure:energy_RMSE}
\end{center}
\end{figure}

\begin{figure}[htb]
\begin{center}
\includegraphics[width=\columnwidth]{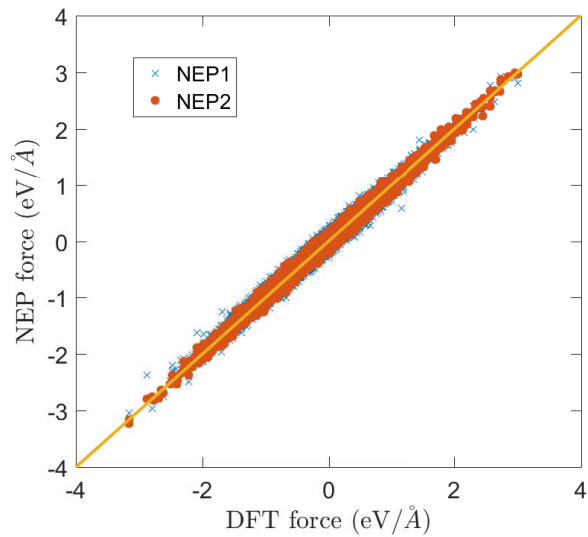}
\caption{Force as calculated from NEP1 and NEP2 for bulk PbTe compared with the training data from quantum mechanical DFT calculations. The solid line represents the identity function used to guide the eyes.}
\label{figure:force_RMSE}
\end{center}
\end{figure}

Figure \ref{figure:energy_RMSE} compares the predicted energies by NEP1 and NEP2 and those from quantum mechanical density functional theory (DFT) calculations. Figure \ref{figure:force_RMSE} shows similar results for force. It can be seen that the energy and force errors from NEP2 are indeed smaller than those from NEP1. Particularly, in NEP1, there are some force errors larger than $0.5$ eV/\AA, which are absent from NEP2.

\begin{figure}[htb]
\begin{center}
\includegraphics[width=\columnwidth]{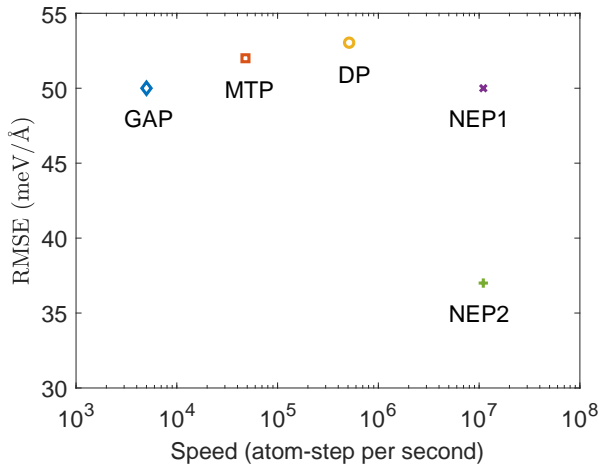}
\caption{Force RMSE and MD speed for the various ML potentials trained using the same set of training data for PbTe. The results for NEP1, GAP, MTP and DP are taken from Ref. \onlinecite{fan2021neuroevolution}. For GAP and MTP, 72 Intel Xeon-Gold 6240 CPU cores are used; for DP and NEP (both NEP1 and NEP2), one Nvidia Tesla V100 GPU card is used. These CPU and GPU resources are of comparable price.}
\label{figure:accuracy_vs_speed}
\end{center}
\end{figure}

To better appreciate the performance of NEP2, we compare it with both NEP1 and some other popular ML potential packages \cite{quip,Novikov2021,wang2018cpc}. Figure \ref{figure:accuracy_vs_speed} shows the force regression accuracy and MD speed for NEP1, NEP2, GAP \cite{quip}, MTP \cite{Novikov2021}, and DP (deep potential)  \cite{wang2018cpc}. The MD speed is measured as the product of the number of atoms and the number of steps that can be achieved per second. We see that NEP1 is already of comparable accuracy to the other ML potentials in this case, and is one to three orders of magnitude faster. NEP2 can achieve a higher accuracy than NEP1 while keeping the speed of NEP1. Therefore, NEP2 can achieve a high accuracy and a high computational speed simultaneously.

\begin{figure}[htb]
\begin{center}
\includegraphics[width=\columnwidth]{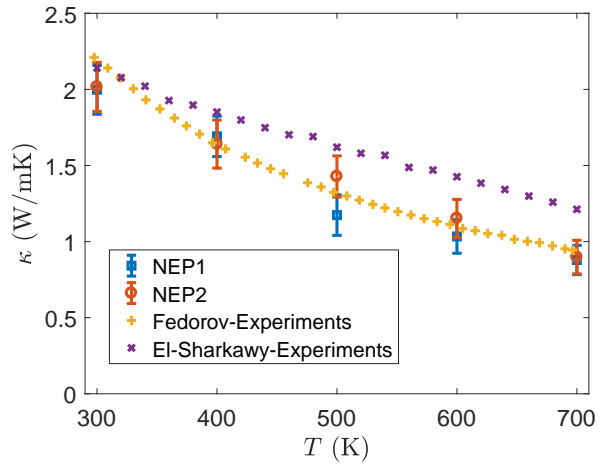}
\caption{Lattice thermal conductivity of bulk PbTe as a function of temperature from HNEMD simulations with NEP1 and NEP2 and from experiments \cite{Fedorov1969,El-Sharkawy1983ijt}. The NEP1 results are taken from Ref. \onlinecite{fan2021neuroevolution}.}
\label{figure:kappa}
\end{center}
\end{figure}

To make sure that the high accuracy of NEP2 is not a result of overfitting, we perform MD simulations to calculate the thermal conductivity of PbTe from 300 K to 700 K. We use the efficient homogeneous nonequilibrium MD (HNEMD) method \cite{Fan2019prb,fan2021neuroevolution} with a driving force parameter of $1.0$ $\mu$m$^{-1}$. We use a cubic simulation cell of $8000$ atoms and a time step of $1.0$ fs. For each temperature, three independent HNEMD simulations are performed, each with a production time of $2000$ ps. In Fig. \ref{figure:kappa}, we compare the thermal conductivity values calculated from NEP1 and NEP2 and experimental ones \cite{Fedorov1969,El-Sharkawy1983ijt}. While the two sets of experimental data have some discrepancies, it can be seen that the lattice thermal conductivity values calculated using NEP2 do not give a worse agreement with the experimental ones than NEP1. This indicates that the high accuracy of NEP2 is not a result of overfitting.

\subsection{Al-Cu-Mg alloy}

\begin{figure}[htb]
\begin{center}
\includegraphics[width=\columnwidth]{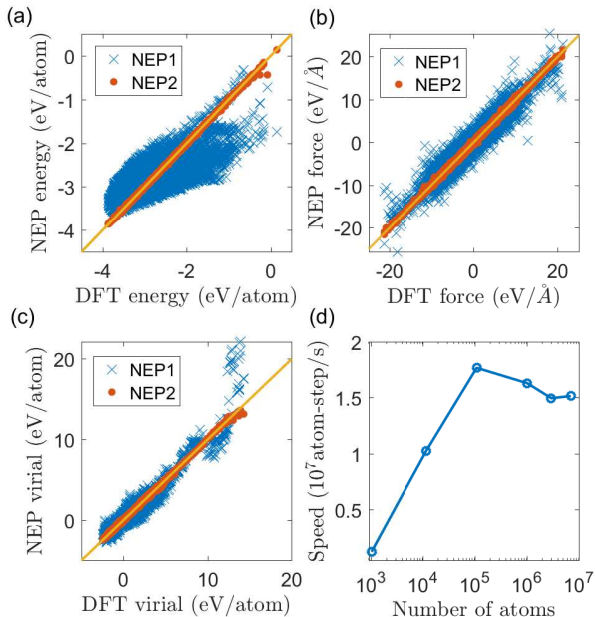}
\caption{(a) Energy, (b) force, and (c) virial as calculated from NEP1 and NEP2 for the Al-Cu-Mg testing data set compared with the DFT training data. The solid lines represent the identity function used to guide the eyes. (d) Computational speed of the NEP2 potential as a function of the number of atoms using one Tesla A100 GPU card. }
\label{figure:MgAlCu}
\end{center}
\end{figure}

We now move from the relatively simple case of bulk PbTe to the more challenging case of Al-Cu-Mg alloy as studied by Jiang \textit{et al}. \cite{jiang2021cpb} using the DeePMD-kit package \cite{wang2018cpc}. A large data set with a full range of relative concentrations of the three atom types has been generated using a concurrent-learning scheme \cite{jiang2021cpb}. There are $141409$ structures with more than three million atoms in total. We randomly select $10^4$ structures as the testing set and use the remaining structures for training. The hypyerparameters we use for NEP1 and NEP2 are listed in Table \ref{table:nep-hyper}.

Figures \ref{figure:MgAlCu}(a)-(c) show the energy, force, and virial as calculated from NEP1 and NEP2 compared with the DFT training data. We see that NEP1 has very large errors in all the quantities, while NEP2 has much higher accuracy (See Table \ref{table:error} for the RMSE and MAE values). Figure \ref{figure:MgAlCu}(d) shows the MD simulation speed of NEP2 as a function of the number of atoms in the simulated system (NEP1 is unstable and we thus have not run MD simulations with it). Using a single Tesla A100 GPU (with $80$ GB device memory), we can run MD simulations with systems containing up to about $7$ million atoms and the computational speed is over $1.5 \times 10^7$ atom-step/second. As a reference, we note that the Al-Cu-Mg DP potential after model compression \cite{lu2021dp} can run MD simulations with systems containing up to about $60$ thousand atoms with a computational speed of about $1.9\times 10^5$ atom-step/second using one Tesla V100 GPU (with 32 GB device memory).

\subsection{Origin of the higher accuracy of NEP2 as compared to NEP1}

\begin{figure}[htb]
\begin{center}
\includegraphics[width=\columnwidth]{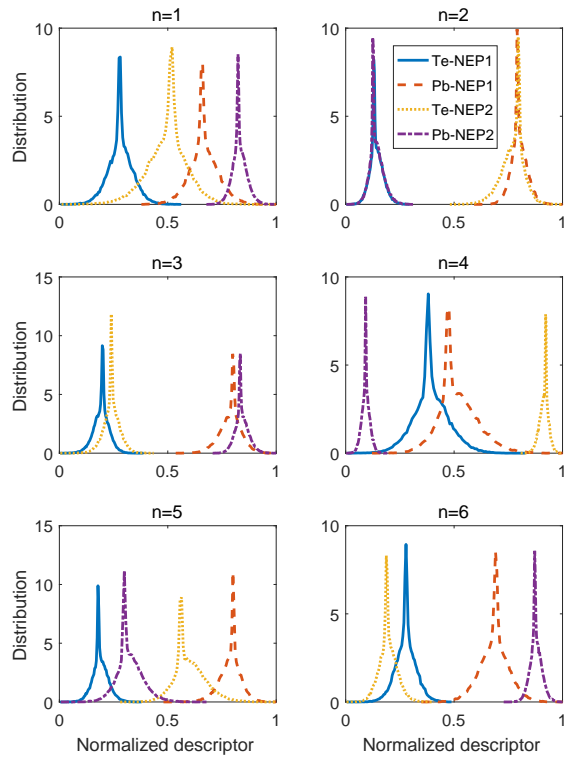}
\caption{Distribution of the normalized radial descriptor components $\{q_{n}\}_{n=1}^{6}$ for the Pb and Te atoms in NEP1 and NEP2.}
\label{figure:descriptor_radial}
\end{center}
\end{figure}

\begin{figure}[htb]
\begin{center}
\includegraphics[width=\columnwidth]{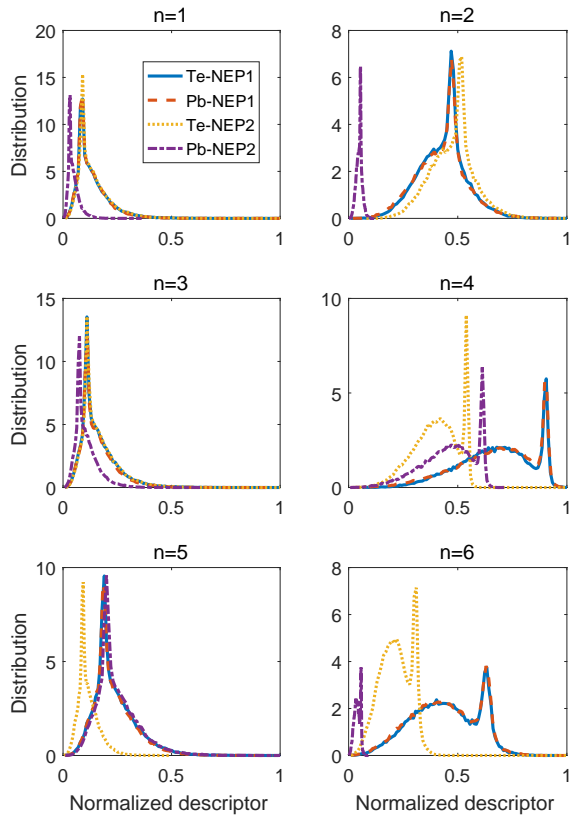}
\caption{Distribution of the normalized angular descriptor components $\{q_{n4}\}_{n=1}^{6}$ (that is, most of the $l=4$ components) for the Pb and Te atoms in NEP1 and NEP2.}
\label{figure:descriptor_angular}
\end{center}
\end{figure}

To understand the origin of the higher accuracy achievable by NEP2 as compared to NEP1, we examine the distributions of some descriptor components in Fig. \ref{figure:descriptor_radial} and Fig. \ref{figure:descriptor_angular}. For the radial components ($q_{n}$ with $n=1$ to $n=6$) shown in Fig. \ref{figure:descriptor_radial}, the distributions for Pb and Te atoms are well distinguishable in both NEP1 and NEP2. However, for the angular components ($q_{n4}$ with $n=1$ to $n=6$) shown in Fig. \ref{figure:descriptor_angular}, the distributions for Pb and Te atoms are almost identical in NEP1 but are well distinguishable in NEP2. The fact that the angular descriptor distributions for Pb and Te atoms are almost identical in NEP1 is related to the choice of $c_{ij}=\sqrt{z_iz_j}$ in Eq. (\ref{equation:g_n}) and the relatively small cutoff for the angular components. In NEP2, without increasing the cutoff for the angular components but simply optimizing the $c_{nij}$ parameters in Eq. (\ref{equation:g_n_nep2}) for each radial function $g_n$ can lead to more distinguishable descriptor distributions for different atom types hence better discrimination of the different atom types in a multi-component system. This is the origin of the higher accuracy of NEP2 as compared to NEP1.

\section{Summary and conclusions \label{section:summary}}

In summary, we have proposed an improved scheme of considering different atom types in the atom-environment descriptor used in the neuroevolution machine-learning potential. The improved method leads to higher regression accuracy without increasing the computational cost in molecular dynamics simulations, as demonstrated using two case studies: bulk PbTe and Al-Cu-Mg alloy. The increased regression accuracy is shown to originate from the improved discrimination of the descriptor distributions for the different atom types in the multi-component system. This improved neuroevolution machine-learning potential is implemented in version $2.9$ of the open-source \textsc{gpumd} package.

\begin{acknowledgments}
We thank Haikuan Dong, Jiahui Liu, Keke Song, Yanzhou Wang, Ke Xu, Penghua Ying, Zezhu Zeng, and many other GPUMD users for testing the versions developed during the course of this research.
ZF acknowledges the supports from the National Natural Science Foundation of China (NSFC) (No. 11974059) and the Science Foundation from Education Department of Liaoning Province under Grant No. LQ2019010. 
\end{acknowledgments}


\begin{thebibliography}{35}%
\makeatletter
\providecommand \@ifxundefined [1]{%
 \@ifx{#1\undefined}
}%
\providecommand \@ifnum [1]{%
 \ifnum #1\expandafter \@firstoftwo
 \else \expandafter \@secondoftwo
 \fi
}%
\providecommand \@ifx [1]{%
 \ifx #1\expandafter \@firstoftwo
 \else \expandafter \@secondoftwo
 \fi
}%
\providecommand \natexlab [1]{#1}%
\providecommand \enquote  [1]{``#1''}%
\providecommand \bibnamefont  [1]{#1}%
\providecommand \bibfnamefont [1]{#1}%
\providecommand \citenamefont [1]{#1}%
\providecommand \href@noop [0]{\@secondoftwo}%
\providecommand \href [0]{\begingroup \@sanitize@url \@href}%
\providecommand \@href[1]{\@@startlink{#1}\@@href}%
\providecommand \@@href[1]{\endgroup#1\@@endlink}%
\providecommand \@sanitize@url [0]{\catcode `\\12\catcode `\$12\catcode
  `\&12\catcode `\#12\catcode `\^12\catcode `\_12\catcode `\%12\relax}%
\providecommand \@@startlink[1]{}%
\providecommand \@@endlink[0]{}%
\providecommand \url  [0]{\begingroup\@sanitize@url \@url }%
\providecommand \@url [1]{\endgroup\@href {#1}{\urlprefix }}%
\providecommand \urlprefix  [0]{URL }%
\providecommand \Eprint [0]{\href }%
\providecommand \doibase [0]{http://dx.doi.org/}%
\providecommand \selectlanguage [0]{\@gobble}%
\providecommand \bibinfo  [0]{\@secondoftwo}%
\providecommand \bibfield  [0]{\@secondoftwo}%
\providecommand \translation [1]{[#1]}%
\providecommand \BibitemOpen [0]{}%
\providecommand \bibitemStop [0]{}%
\providecommand \bibitemNoStop [0]{.\EOS\space}%
\providecommand \EOS [0]{\spacefactor3000\relax}%
\providecommand \BibitemShut  [1]{\csname bibitem#1\endcsname}%
\let\auto@bib@innerbib\@empty
\bibitem [{\citenamefont {Behler}(2016)}]{behler2016jcp}%
  \BibitemOpen
  \bibfield  {author} {\bibinfo {author} {\bibfnamefont {J\"org}\ \bibnamefont
  {Behler}},\ }\bibfield  {title} {\enquote {\bibinfo {title} {Perspective:
  Machine learning potentials for atomistic simulations},}\ }\href {\doibase
  10.1063/1.4966192} {\bibfield  {journal} {\bibinfo  {journal} {The Journal of
  Chemical Physics}\ }\textbf {\bibinfo {volume} {145}},\ \bibinfo {pages}
  {170901} (\bibinfo {year} {2016})}\BibitemShut {NoStop}%
\bibitem [{\citenamefont {Deringer}\ \emph {et~al.}(2019)\citenamefont
  {Deringer}, \citenamefont {Caro},\ and\ \citenamefont
  {Cs\'anyi}}]{Deringer2019am}%
  \BibitemOpen
  \bibfield  {author} {\bibinfo {author} {\bibfnamefont {Volker~L.}\
  \bibnamefont {Deringer}}, \bibinfo {author} {\bibfnamefont {Miguel~A.}\
  \bibnamefont {Caro}}, \ and\ \bibinfo {author} {\bibfnamefont {G\'abor}\
  \bibnamefont {Cs\'anyi}},\ }\bibfield  {title} {\enquote {\bibinfo {title}
  {Machine learning interatomic potentials as emerging tools for materials
  science},}\ }\href {\doibase https://doi.org/10.1002/adma.201902765}
  {\bibfield  {journal} {\bibinfo  {journal} {Advanced Materials}\ }\textbf
  {\bibinfo {volume} {31}},\ \bibinfo {pages} {1902765} (\bibinfo {year}
  {2019})}\BibitemShut {NoStop}%
\bibitem [{\citenamefont {Mueller}\ \emph {et~al.}(2020)\citenamefont
  {Mueller}, \citenamefont {Hernandez},\ and\ \citenamefont
  {Wang}}]{Mueller2020jcp}%
  \BibitemOpen
  \bibfield  {author} {\bibinfo {author} {\bibfnamefont {Tim}\ \bibnamefont
  {Mueller}}, \bibinfo {author} {\bibfnamefont {Alberto}\ \bibnamefont
  {Hernandez}}, \ and\ \bibinfo {author} {\bibfnamefont {Chuhong}\ \bibnamefont
  {Wang}},\ }\bibfield  {title} {\enquote {\bibinfo {title} {Machine learning
  for interatomic potential models},}\ }\href {\doibase 10.1063/1.5126336}
  {\bibfield  {journal} {\bibinfo  {journal} {The Journal of Chemical Physics}\
  }\textbf {\bibinfo {volume} {152}},\ \bibinfo {pages} {050902} (\bibinfo
  {year} {2020})}\BibitemShut {NoStop}%
\bibitem [{\citenamefont {Mishin}(2021)}]{Mishin2021am}%
  \BibitemOpen
  \bibfield  {author} {\bibinfo {author} {\bibfnamefont {Y.}~\bibnamefont
  {Mishin}},\ }\bibfield  {title} {\enquote {\bibinfo {title} {Machine-learning
  interatomic potentials for materials science},}\ }\href {\doibase
  https://doi.org/10.1016/j.actamat.2021.116980} {\bibfield  {journal}
  {\bibinfo  {journal} {Acta Materialia}\ }\textbf {\bibinfo {volume} {214}},\
  \bibinfo {pages} {116980} (\bibinfo {year} {2021})}\BibitemShut {NoStop}%
\bibitem [{\citenamefont {Unke}\ \emph {et~al.}(2021)\citenamefont {Unke},
  \citenamefont {Chmiela}, \citenamefont {Sauceda}, \citenamefont {Gastegger},
  \citenamefont {Poltavsky}, \citenamefont {Schütt}, \citenamefont
  {Tkatchenko},\ and\ \citenamefont {Müller}}]{Unke2021cr}%
  \BibitemOpen
  \bibfield  {author} {\bibinfo {author} {\bibfnamefont {Oliver~T.}\
  \bibnamefont {Unke}}, \bibinfo {author} {\bibfnamefont {Stefan}\ \bibnamefont
  {Chmiela}}, \bibinfo {author} {\bibfnamefont {Huziel~E.}\ \bibnamefont
  {Sauceda}}, \bibinfo {author} {\bibfnamefont {Michael}\ \bibnamefont
  {Gastegger}}, \bibinfo {author} {\bibfnamefont {Igor}\ \bibnamefont
  {Poltavsky}}, \bibinfo {author} {\bibfnamefont {Kristof~T.}\ \bibnamefont
  {Schütt}}, \bibinfo {author} {\bibfnamefont {Alexandre}\ \bibnamefont
  {Tkatchenko}}, \ and\ \bibinfo {author} {\bibfnamefont {Klaus-Robert}\
  \bibnamefont {Müller}},\ }\bibfield  {title} {\enquote {\bibinfo {title}
  {Machine learning force fields},}\ }\href {\doibase
  10.1021/acs.chemrev.0c01111} {\bibfield  {journal} {\bibinfo  {journal}
  {Chemical Reviews}\ }\textbf {\bibinfo {volume} {121}},\ \bibinfo {pages}
  {10142--10186} (\bibinfo {year} {2021})}\BibitemShut {NoStop}%
\bibitem [{\citenamefont {Behler}\ and\ \citenamefont
  {Parrinello}(2007)}]{behler2007prl}%
  \BibitemOpen
  \bibfield  {author} {\bibinfo {author} {\bibfnamefont {J\"org}\ \bibnamefont
  {Behler}}\ and\ \bibinfo {author} {\bibfnamefont {Michele}\ \bibnamefont
  {Parrinello}},\ }\bibfield  {title} {\enquote {\bibinfo {title} {Generalized
  neural-network representation of high-dimensional potential-energy
  surfaces},}\ }\href {\doibase 10.1103/PhysRevLett.98.146401} {\bibfield
  {journal} {\bibinfo  {journal} {Phys. Rev. Lett.}\ }\textbf {\bibinfo
  {volume} {98}},\ \bibinfo {pages} {146401} (\bibinfo {year}
  {2007})}\BibitemShut {NoStop}%
\bibitem [{\citenamefont {Bart\'ok}\ \emph {et~al.}(2010)\citenamefont
  {Bart\'ok}, \citenamefont {Payne}, \citenamefont {Kondor},\ and\
  \citenamefont {Cs\'anyi}}]{bartok2010prl}%
  \BibitemOpen
  \bibfield  {author} {\bibinfo {author} {\bibfnamefont {Albert~P.}\
  \bibnamefont {Bart\'ok}}, \bibinfo {author} {\bibfnamefont {Mike~C.}\
  \bibnamefont {Payne}}, \bibinfo {author} {\bibfnamefont {Risi}\ \bibnamefont
  {Kondor}}, \ and\ \bibinfo {author} {\bibfnamefont {G\'abor}\ \bibnamefont
  {Cs\'anyi}},\ }\bibfield  {title} {\enquote {\bibinfo {title} {{Gaussian
  Approximation Potentials: The Accuracy of Quantum Mechanics, without the
  Electrons}},}\ }\href {\doibase 10.1103/PhysRevLett.104.136403} {\bibfield
  {journal} {\bibinfo  {journal} {Phys. Rev. Lett.}\ }\textbf {\bibinfo
  {volume} {104}},\ \bibinfo {pages} {136403} (\bibinfo {year}
  {2010})}\BibitemShut {NoStop}%
\bibitem [{\citenamefont {Thompson}\ \emph {et~al.}(2015)\citenamefont
  {Thompson}, \citenamefont {Swiler}, \citenamefont {Trott}, \citenamefont
  {Foiles},\ and\ \citenamefont {Tucker}}]{Thompson2014jcp}%
  \BibitemOpen
  \bibfield  {author} {\bibinfo {author} {\bibfnamefont {A.P.}\ \bibnamefont
  {Thompson}}, \bibinfo {author} {\bibfnamefont {L.P.}\ \bibnamefont {Swiler}},
  \bibinfo {author} {\bibfnamefont {C.R.}\ \bibnamefont {Trott}}, \bibinfo
  {author} {\bibfnamefont {S.M.}\ \bibnamefont {Foiles}}, \ and\ \bibinfo
  {author} {\bibfnamefont {G.J.}\ \bibnamefont {Tucker}},\ }\bibfield  {title}
  {\enquote {\bibinfo {title} {Spectral neighbor analysis method for automated
  generation of quantum-accurate interatomic potentials},}\ }\href {\doibase
  https://doi.org/10.1016/j.jcp.2014.12.018} {\bibfield  {journal} {\bibinfo
  {journal} {Journal of Computational Physics}\ }\textbf {\bibinfo {volume}
  {285}},\ \bibinfo {pages} {316--330} (\bibinfo {year} {2015})}\BibitemShut
  {NoStop}%
\bibitem [{\citenamefont {Shapeev}(2016)}]{Shapeev2016}%
  \BibitemOpen
  \bibfield  {author} {\bibinfo {author} {\bibfnamefont {Alexander~V.}\
  \bibnamefont {Shapeev}},\ }\bibfield  {title} {\enquote {\bibinfo {title}
  {Moment tensor potentials: A class of systematically improvable interatomic
  potentials},}\ }\href {\doibase 10.1137/15M1054183} {\bibfield  {journal}
  {\bibinfo  {journal} {Multiscale Modeling \& Simulation}\ }\textbf {\bibinfo
  {volume} {14}},\ \bibinfo {pages} {1153--1173} (\bibinfo {year}
  {2016})}\BibitemShut {NoStop}%
\bibitem [{\citenamefont {Wang}\ \emph {et~al.}(2018)\citenamefont {Wang},
  \citenamefont {Zhang}, \citenamefont {Han},\ and\ \citenamefont
  {E}}]{wang2018cpc}%
  \BibitemOpen
  \bibfield  {author} {\bibinfo {author} {\bibfnamefont {Han}\ \bibnamefont
  {Wang}}, \bibinfo {author} {\bibfnamefont {Linfeng}\ \bibnamefont {Zhang}},
  \bibinfo {author} {\bibfnamefont {Jiequn}\ \bibnamefont {Han}}, \ and\
  \bibinfo {author} {\bibfnamefont {Weinan}\ \bibnamefont {E}},\ }\bibfield
  {title} {\enquote {\bibinfo {title} {{DeePMD-kit: A deep learning package for
  many-body potential energy representation and molecular dynamics}},}\ }\href
  {\doibase https://doi.org/10.1016/j.cpc.2018.03.016} {\bibfield  {journal}
  {\bibinfo  {journal} {Computer Physics Communications}\ }\textbf {\bibinfo
  {volume} {228}},\ \bibinfo {pages} {178--184} (\bibinfo {year}
  {2018})}\BibitemShut {NoStop}%
\bibitem [{\citenamefont {Zhang}\ \emph {et~al.}(2018)\citenamefont {Zhang},
  \citenamefont {Han}, \citenamefont {Wang}, \citenamefont {Car},\ and\
  \citenamefont {E}}]{zhang2018prl}%
  \BibitemOpen
  \bibfield  {author} {\bibinfo {author} {\bibfnamefont {Linfeng}\ \bibnamefont
  {Zhang}}, \bibinfo {author} {\bibfnamefont {Jiequn}\ \bibnamefont {Han}},
  \bibinfo {author} {\bibfnamefont {Han}\ \bibnamefont {Wang}}, \bibinfo
  {author} {\bibfnamefont {Roberto}\ \bibnamefont {Car}}, \ and\ \bibinfo
  {author} {\bibfnamefont {Weinan}\ \bibnamefont {E}},\ }\bibfield  {title}
  {\enquote {\bibinfo {title} {Deep potential molecular dynamics: A scalable
  model with the accuracy of quantum mechanics},}\ }\href {\doibase
  10.1103/PhysRevLett.120.143001} {\bibfield  {journal} {\bibinfo  {journal}
  {Phys. Rev. Lett.}\ }\textbf {\bibinfo {volume} {120}},\ \bibinfo {pages}
  {143001} (\bibinfo {year} {2018})}\BibitemShut {NoStop}%
\bibitem [{\citenamefont {Lee}\ \emph {et~al.}(2019)\citenamefont {Lee},
  \citenamefont {Yoo}, \citenamefont {Jeong},\ and\ \citenamefont
  {Han}}]{lee2019cpc}%
  \BibitemOpen
  \bibfield  {author} {\bibinfo {author} {\bibfnamefont {Kyuhyun}\ \bibnamefont
  {Lee}}, \bibinfo {author} {\bibfnamefont {Dongsun}\ \bibnamefont {Yoo}},
  \bibinfo {author} {\bibfnamefont {Wonseok}\ \bibnamefont {Jeong}}, \ and\
  \bibinfo {author} {\bibfnamefont {Seungwu}\ \bibnamefont {Han}},\ }\bibfield
  {title} {\enquote {\bibinfo {title} {{SIMPLE-NN: An efficient package for
  training and executing neural-network interatomic potentials}},}\ }\href
  {\doibase https://doi.org/10.1016/j.cpc.2019.04.014} {\bibfield  {journal}
  {\bibinfo  {journal} {Computer Physics Communications}\ }\textbf {\bibinfo
  {volume} {242}},\ \bibinfo {pages} {95--103} (\bibinfo {year}
  {2019})}\BibitemShut {NoStop}%
\bibitem [{\citenamefont {Lot}\ \emph {et~al.}(2020)\citenamefont {Lot},
  \citenamefont {Pellegrini}, \citenamefont {Shaidu},\ and\ \citenamefont
  {Küçükbenli}}]{lot2020cpc}%
  \BibitemOpen
  \bibfield  {author} {\bibinfo {author} {\bibfnamefont {Ruggero}\ \bibnamefont
  {Lot}}, \bibinfo {author} {\bibfnamefont {Franco}\ \bibnamefont
  {Pellegrini}}, \bibinfo {author} {\bibfnamefont {Yusuf}\ \bibnamefont
  {Shaidu}}, \ and\ \bibinfo {author} {\bibfnamefont {Emine}\ \bibnamefont
  {Küçükbenli}},\ }\bibfield  {title} {\enquote {\bibinfo {title} {{PANNA:
  Properties from Artificial Neural Network Architectures}},}\ }\href {\doibase
  https://doi.org/10.1016/j.cpc.2020.107402} {\bibfield  {journal} {\bibinfo
  {journal} {Computer Physics Communications}\ }\textbf {\bibinfo {volume}
  {256}},\ \bibinfo {pages} {107402} (\bibinfo {year} {2020})}\BibitemShut
  {NoStop}%
\bibitem [{\citenamefont {Gao}\ \emph {et~al.}(2020)\citenamefont {Gao},
  \citenamefont {Ramezanghorbani}, \citenamefont {Isayev}, \citenamefont
  {Smith},\ and\ \citenamefont {Roitberg}}]{gao2020jcim}%
  \BibitemOpen
  \bibfield  {author} {\bibinfo {author} {\bibfnamefont {Xiang}\ \bibnamefont
  {Gao}}, \bibinfo {author} {\bibfnamefont {Farhad}\ \bibnamefont
  {Ramezanghorbani}}, \bibinfo {author} {\bibfnamefont {Olexandr}\ \bibnamefont
  {Isayev}}, \bibinfo {author} {\bibfnamefont {Justin~S.}\ \bibnamefont
  {Smith}}, \ and\ \bibinfo {author} {\bibfnamefont {Adrian~E.}\ \bibnamefont
  {Roitberg}},\ }\bibfield  {title} {\enquote {\bibinfo {title} {{TorchANI: A
  Free and Open Source PyTorch-Based Deep Learning Implementation of the ANI
  Neural Network Potentials}},}\ }\href {\doibase 10.1021/acs.jcim.0c00451}
  {\bibfield  {journal} {\bibinfo  {journal} {Journal of Chemical Information
  and Modeling}\ }\textbf {\bibinfo {volume} {60}},\ \bibinfo {pages}
  {3408--3415} (\bibinfo {year} {2020})}\BibitemShut {NoStop}%
\bibitem [{\citenamefont {Shao}\ \emph {et~al.}(2020)\citenamefont {Shao},
  \citenamefont {Hellström}, \citenamefont {Mitev}, \citenamefont {Knijff},\
  and\ \citenamefont {Zhang}}]{shao2020jcim}%
  \BibitemOpen
  \bibfield  {author} {\bibinfo {author} {\bibfnamefont {Yunqi}\ \bibnamefont
  {Shao}}, \bibinfo {author} {\bibfnamefont {Matti}\ \bibnamefont
  {Hellström}}, \bibinfo {author} {\bibfnamefont {Pavlin~D.}\ \bibnamefont
  {Mitev}}, \bibinfo {author} {\bibfnamefont {Lisanne}\ \bibnamefont {Knijff}},
  \ and\ \bibinfo {author} {\bibfnamefont {Chao}\ \bibnamefont {Zhang}},\
  }\bibfield  {title} {\enquote {\bibinfo {title} {{PiNN: A Python Library for
  Building Atomic Neural Networks of Molecules and Materials}},}\ }\href
  {\doibase 10.1021/acs.jcim.9b00994} {\bibfield  {journal} {\bibinfo
  {journal} {Journal of Chemical Information and Modeling}\ }\textbf {\bibinfo
  {volume} {60}},\ \bibinfo {pages} {1184--1193} (\bibinfo {year}
  {2020})}\BibitemShut {NoStop}%
\bibitem [{\citenamefont {Pattnaik}\ \emph {et~al.}(2020)\citenamefont
  {Pattnaik}, \citenamefont {Raghunathan}, \citenamefont {Kalluri},
  \citenamefont {Bhimalapuram}, \citenamefont {Jawahar},\ and\ \citenamefont
  {Priyakumar}}]{Pattnaik2020jpca}%
  \BibitemOpen
  \bibfield  {author} {\bibinfo {author} {\bibfnamefont {Punyaslok}\
  \bibnamefont {Pattnaik}}, \bibinfo {author} {\bibfnamefont {Shampa}\
  \bibnamefont {Raghunathan}}, \bibinfo {author} {\bibfnamefont {Tarun}\
  \bibnamefont {Kalluri}}, \bibinfo {author} {\bibfnamefont {Prabhakar}\
  \bibnamefont {Bhimalapuram}}, \bibinfo {author} {\bibfnamefont {C.~V.}\
  \bibnamefont {Jawahar}}, \ and\ \bibinfo {author} {\bibfnamefont {U.~Deva}\
  \bibnamefont {Priyakumar}},\ }\bibfield  {title} {\enquote {\bibinfo {title}
  {{Machine Learning for Accurate Force Calculations in Molecular Dynamics
  Simulations}},}\ }\href {\doibase 10.1021/acs.jpca.0c03926} {\bibfield
  {journal} {\bibinfo  {journal} {The Journal of Physical Chemistry A}\
  }\textbf {\bibinfo {volume} {124}},\ \bibinfo {pages} {6954--6967} (\bibinfo
  {year} {2020})}\BibitemShut {NoStop}%
\bibitem [{\citenamefont {Yanxon}\ \emph {et~al.}(2021)\citenamefont {Yanxon},
  \citenamefont {Zagaceta}, \citenamefont {Tang}, \citenamefont {Matteson},\
  and\ \citenamefont {Zhu}}]{Yanxon2021}%
  \BibitemOpen
  \bibfield  {author} {\bibinfo {author} {\bibfnamefont {Howard}\ \bibnamefont
  {Yanxon}}, \bibinfo {author} {\bibfnamefont {David}\ \bibnamefont
  {Zagaceta}}, \bibinfo {author} {\bibfnamefont {Binh}\ \bibnamefont {Tang}},
  \bibinfo {author} {\bibfnamefont {David~S}\ \bibnamefont {Matteson}}, \ and\
  \bibinfo {author} {\bibfnamefont {Qiang}\ \bibnamefont {Zhu}},\ }\bibfield
  {title} {\enquote {\bibinfo {title} {{PyXtal}{\_}{FF}: a python library for
  automated force field generation},}\ }\href {\doibase
  10.1088/2632-2153/abc940} {\bibfield  {journal} {\bibinfo  {journal} {Machine
  Learning: Science and Technology}\ }\textbf {\bibinfo {volume} {2}},\
  \bibinfo {pages} {027001} (\bibinfo {year} {2021})}\BibitemShut {NoStop}%
\bibitem [{\citenamefont {Zhang}\ \emph {et~al.}(2021)\citenamefont {Zhang},
  \citenamefont {Xia},\ and\ \citenamefont {Jiang}}]{zhang2021prl}%
  \BibitemOpen
  \bibfield  {author} {\bibinfo {author} {\bibfnamefont {Yaolong}\ \bibnamefont
  {Zhang}}, \bibinfo {author} {\bibfnamefont {Junfan}\ \bibnamefont {Xia}}, \
  and\ \bibinfo {author} {\bibfnamefont {Bin}\ \bibnamefont {Jiang}},\
  }\bibfield  {title} {\enquote {\bibinfo {title} {{Physically Motivated
  Recursively Embedded Atom Neural Networks: Incorporating Local Completeness
  and Nonlocality}},}\ }\href {\doibase 10.1103/PhysRevLett.127.156002}
  {\bibfield  {journal} {\bibinfo  {journal} {Phys. Rev. Lett.}\ }\textbf
  {\bibinfo {volume} {127}},\ \bibinfo {pages} {156002} (\bibinfo {year}
  {2021})}\BibitemShut {NoStop}%
\bibitem [{\citenamefont {Fan}\ \emph {et~al.}(2021)\citenamefont {Fan},
  \citenamefont {Zeng}, \citenamefont {Zhang}, \citenamefont {Wang},
  \citenamefont {Song}, \citenamefont {Dong}, \citenamefont {Chen},\ and\
  \citenamefont {Ala-Nissila}}]{fan2021neuroevolution}%
  \BibitemOpen
  \bibfield  {author} {\bibinfo {author} {\bibfnamefont {Zheyong}\ \bibnamefont
  {Fan}}, \bibinfo {author} {\bibfnamefont {Zezhu}\ \bibnamefont {Zeng}},
  \bibinfo {author} {\bibfnamefont {Cunzhi}\ \bibnamefont {Zhang}}, \bibinfo
  {author} {\bibfnamefont {Yanzhou}\ \bibnamefont {Wang}}, \bibinfo {author}
  {\bibfnamefont {Keke}\ \bibnamefont {Song}}, \bibinfo {author} {\bibfnamefont
  {Haikuan}\ \bibnamefont {Dong}}, \bibinfo {author} {\bibfnamefont {Yue}\
  \bibnamefont {Chen}}, \ and\ \bibinfo {author} {\bibfnamefont {Tapio}\
  \bibnamefont {Ala-Nissila}},\ }\bibfield  {title} {\enquote {\bibinfo {title}
  {Neuroevolution machine learning potentials: Combining high accuracy and low
  cost in atomistic simulations and application to heat transport},}\ }\href
  {\doibase 10.1103/PhysRevB.104.104309} {\bibfield  {journal} {\bibinfo
  {journal} {Phys. Rev. B}\ }\textbf {\bibinfo {volume} {104}},\ \bibinfo
  {pages} {104309} (\bibinfo {year} {2021})}\BibitemShut {NoStop}%
\bibitem [{\citenamefont {Schaul}\ \emph {et~al.}(2011)\citenamefont {Schaul},
  \citenamefont {Glasmachers},\ and\ \citenamefont {Schmidhuber}}]{Schaul2011}%
  \BibitemOpen
  \bibfield  {author} {\bibinfo {author} {\bibfnamefont {Tom}\ \bibnamefont
  {Schaul}}, \bibinfo {author} {\bibfnamefont {Tobias}\ \bibnamefont
  {Glasmachers}}, \ and\ \bibinfo {author} {\bibfnamefont {J\"{u}rgen}\
  \bibnamefont {Schmidhuber}},\ }\bibfield  {title} {\enquote {\bibinfo {title}
  {High dimensions and heavy tails for natural evolution strategies},}\ }in\
  \href {\doibase 10.1145/2001576.2001692} {\emph {\bibinfo {booktitle}
  {Proceedings of the 13th Annual Conference on Genetic and Evolutionary
  Computation}}},\ \bibinfo {series and number} {GECCO '11}\ (\bibinfo
  {publisher} {Association for Computing Machinery},\ \bibinfo {address} {New
  York, NY, USA},\ \bibinfo {year} {2011})\ pp.\ \bibinfo {pages}
  {845--852}\BibitemShut {NoStop}%
\bibitem [{\citenamefont {Wierstra}\ \emph {et~al.}(2014)\citenamefont
  {Wierstra}, \citenamefont {Schaul}, \citenamefont {Glasmachers},
  \citenamefont {Sun}, \citenamefont {Peters},\ and\ \citenamefont
  {Schmidhuber}}]{wierstra2014jmlr}%
  \BibitemOpen
  \bibfield  {author} {\bibinfo {author} {\bibfnamefont {Daan}\ \bibnamefont
  {Wierstra}}, \bibinfo {author} {\bibfnamefont {Tom}\ \bibnamefont {Schaul}},
  \bibinfo {author} {\bibfnamefont {Tobias}\ \bibnamefont {Glasmachers}},
  \bibinfo {author} {\bibfnamefont {Yi}~\bibnamefont {Sun}}, \bibinfo {author}
  {\bibfnamefont {Jan}\ \bibnamefont {Peters}}, \ and\ \bibinfo {author}
  {\bibfnamefont {J\"{u}rgen}\ \bibnamefont {Schmidhuber}},\ }\bibfield
  {title} {\enquote {\bibinfo {title} {Natural evolution strategies},}\ }\href
  {http://jmlr.org/papers/v15/wierstra14a.html} {\bibfield  {journal} {\bibinfo
   {journal} {Journal of Machine Learning Research}\ }\textbf {\bibinfo
  {volume} {15}},\ \bibinfo {pages} {949--980} (\bibinfo {year}
  {2014})}\BibitemShut {NoStop}%
\bibitem [{\citenamefont {Fan}\ \emph {et~al.}(2013)\citenamefont {Fan},
  \citenamefont {Siro},\ and\ \citenamefont {Harju}}]{fan2013cpc}%
  \BibitemOpen
  \bibfield  {author} {\bibinfo {author} {\bibfnamefont {Zheyong}\ \bibnamefont
  {Fan}}, \bibinfo {author} {\bibfnamefont {Topi}\ \bibnamefont {Siro}}, \ and\
  \bibinfo {author} {\bibfnamefont {Ari}\ \bibnamefont {Harju}},\ }\bibfield
  {title} {\enquote {\bibinfo {title} {Accelerated molecular dynamics force
  evaluation on graphics processing units for thermal conductivity
  calculations},}\ }\href {\doibase
  http://dx.doi.org/10.1016/j.cpc.2013.01.008} {\bibfield  {journal} {\bibinfo
  {journal} {Computer Physics Communications}\ }\textbf {\bibinfo {volume}
  {184}},\ \bibinfo {pages} {1414 -- 1425} (\bibinfo {year}
  {2013})}\BibitemShut {NoStop}%
\bibitem [{\citenamefont {Fan}\ \emph {et~al.}(2017)\citenamefont {Fan},
  \citenamefont {Chen}, \citenamefont {Vierimaa},\ and\ \citenamefont
  {Harju}}]{fan2017cpc}%
  \BibitemOpen
  \bibfield  {author} {\bibinfo {author} {\bibfnamefont {Zheyong}\ \bibnamefont
  {Fan}}, \bibinfo {author} {\bibfnamefont {Wei}\ \bibnamefont {Chen}},
  \bibinfo {author} {\bibfnamefont {Ville}\ \bibnamefont {Vierimaa}}, \ and\
  \bibinfo {author} {\bibfnamefont {Ari}\ \bibnamefont {Harju}},\ }\bibfield
  {title} {\enquote {\bibinfo {title} {Efficient molecular dynamics simulations
  with many-body potentials on graphics processing units},}\ }\href {\doibase
  https://doi.org/10.1016/j.cpc.2017.05.003} {\bibfield  {journal} {\bibinfo
  {journal} {Computer Physics Communications}\ }\textbf {\bibinfo {volume}
  {218}},\ \bibinfo {pages} {10 -- 16} (\bibinfo {year} {2017})}\BibitemShut
  {NoStop}%
\bibitem [{gpu()}]{gpumd-github}%
  \BibitemOpen
  \href@noop {} {}\bibinfo {howpublished}
  {\url{https://github.com/brucefan1983/GPUMD}}\BibitemShut {NoStop}%
\bibitem [{qui()}]{quip}%
  \BibitemOpen
  \href@noop {} {}\bibinfo {howpublished}
  {\url{https://github.com/libAtoms/QUIP}}\BibitemShut {NoStop}%
\bibitem [{\citenamefont {Novikov}\ \emph {et~al.}(2021)\citenamefont
  {Novikov}, \citenamefont {Gubaev}, \citenamefont {Podryabinkin},\ and\
  \citenamefont {Shapeev}}]{Novikov2021}%
  \BibitemOpen
  \bibfield  {author} {\bibinfo {author} {\bibfnamefont {Ivan~S}\ \bibnamefont
  {Novikov}}, \bibinfo {author} {\bibfnamefont {Konstantin}\ \bibnamefont
  {Gubaev}}, \bibinfo {author} {\bibfnamefont {Evgeny~V}\ \bibnamefont
  {Podryabinkin}}, \ and\ \bibinfo {author} {\bibfnamefont {Alexander~V}\
  \bibnamefont {Shapeev}},\ }\bibfield  {title} {\enquote {\bibinfo {title}
  {The {MLIP} package: moment tensor potentials with {MPI} and active
  learning},}\ }\href {\doibase 10.1088/2632-2153/abc9fe} {\bibfield  {journal}
  {\bibinfo  {journal} {Machine Learning: Science and Technology}\ }\textbf
  {\bibinfo {volume} {2}},\ \bibinfo {pages} {025002} (\bibinfo {year}
  {2021})}\BibitemShut {NoStop}%
\bibitem [{\citenamefont {Gastegger}\ \emph {et~al.}(2018)\citenamefont
  {Gastegger}, \citenamefont {Schwiedrzik}, \citenamefont {Bittermann},
  \citenamefont {Berzsenyi},\ and\ \citenamefont
  {Marquetand}}]{Gastegger2018jcp}%
  \BibitemOpen
  \bibfield  {author} {\bibinfo {author} {\bibfnamefont {M.}~\bibnamefont
  {Gastegger}}, \bibinfo {author} {\bibfnamefont {L.}~\bibnamefont
  {Schwiedrzik}}, \bibinfo {author} {\bibfnamefont {M.}~\bibnamefont
  {Bittermann}}, \bibinfo {author} {\bibfnamefont {F.}~\bibnamefont
  {Berzsenyi}}, \ and\ \bibinfo {author} {\bibfnamefont {P.}~\bibnamefont
  {Marquetand}},\ }\bibfield  {title} {\enquote {\bibinfo {title}
  {{wACSF--Weighted atom-centered symmetry functions as descriptors in machine
  learning potentials}},}\ }\href {\doibase 10.1063/1.5019667} {\bibfield
  {journal} {\bibinfo  {journal} {The Journal of Chemical Physics}\ }\textbf
  {\bibinfo {volume} {148}},\ \bibinfo {pages} {241709} (\bibinfo {year}
  {2018})}\BibitemShut {NoStop}%
\bibitem [{\citenamefont {Artrith}\ \emph {et~al.}(2017)\citenamefont
  {Artrith}, \citenamefont {Urban},\ and\ \citenamefont
  {Ceder}}]{Artrith2017prb}%
  \BibitemOpen
  \bibfield  {author} {\bibinfo {author} {\bibfnamefont {Nongnuch}\
  \bibnamefont {Artrith}}, \bibinfo {author} {\bibfnamefont {Alexander}\
  \bibnamefont {Urban}}, \ and\ \bibinfo {author} {\bibfnamefont {Gerbrand}\
  \bibnamefont {Ceder}},\ }\bibfield  {title} {\enquote {\bibinfo {title}
  {Efficient and accurate machine-learning interpolation of atomic energies in
  compositions with many species},}\ }\href {\doibase
  10.1103/PhysRevB.96.014112} {\bibfield  {journal} {\bibinfo  {journal} {Phys.
  Rev. B}\ }\textbf {\bibinfo {volume} {96}},\ \bibinfo {pages} {014112}
  (\bibinfo {year} {2017})}\BibitemShut {NoStop}%
\bibitem [{\citenamefont {Gubaev}\ \emph {et~al.}(2019)\citenamefont {Gubaev},
  \citenamefont {Podryabinkin}, \citenamefont {Hart},\ and\ \citenamefont
  {Shapeev}}]{Gubaev2019cms}%
  \BibitemOpen
  \bibfield  {author} {\bibinfo {author} {\bibfnamefont {Konstantin}\
  \bibnamefont {Gubaev}}, \bibinfo {author} {\bibfnamefont {Evgeny~V.}\
  \bibnamefont {Podryabinkin}}, \bibinfo {author} {\bibfnamefont {Gus~L.W.}\
  \bibnamefont {Hart}}, \ and\ \bibinfo {author} {\bibfnamefont {Alexander~V.}\
  \bibnamefont {Shapeev}},\ }\bibfield  {title} {\enquote {\bibinfo {title}
  {{Accelerating high-throughput searches for new alloys with active learning
  of interatomic potentials}},}\ }\href {\doibase
  https://doi.org/10.1016/j.commatsci.2018.09.031} {\bibfield  {journal}
  {\bibinfo  {journal} {Computational Materials Science}\ }\textbf {\bibinfo
  {volume} {156}},\ \bibinfo {pages} {148--156} (\bibinfo {year}
  {2019})}\BibitemShut {NoStop}%
\bibitem [{nep()}]{nep-data}%
  \BibitemOpen
  \href@noop {} {}\bibinfo {howpublished}
  {\url{https://gitlab.com/brucefan1983/nep-data}}\BibitemShut {NoStop}%
\bibitem [{\citenamefont {Fedorov}\ and\ \citenamefont
  {Machuev}(1969)}]{Fedorov1969}%
  \BibitemOpen
  \bibfield  {author} {\bibinfo {author} {\bibfnamefont {V.I.}\ \bibnamefont
  {Fedorov}}\ and\ \bibinfo {author} {\bibfnamefont {V.I.}\ \bibnamefont
  {Machuev}},\ }\bibfield  {title} {\enquote {\bibinfo {title} {{Thermal
  Conductivity of PbTe, SnTe and GeTe in the solid and liquid phases}},}\
  }\href@noop {} {\bibfield  {journal} {\bibinfo  {journal} {Sov. Phys. Solid
  State USSR}\ }\textbf {\bibinfo {volume} {11}},\ \bibinfo {pages} {1116}
  (\bibinfo {year} {1969})}\BibitemShut {NoStop}%
\bibitem [{\citenamefont {El-Sharkawy}\ \emph {et~al.}(1983)\citenamefont
  {El-Sharkawy}, \citenamefont {Abou El-Azm}, \citenamefont {Kenawy},
  \citenamefont {Hillal},\ and\ \citenamefont
  {Abu-Basha}}]{El-Sharkawy1983ijt}%
  \BibitemOpen
  \bibfield  {author} {\bibinfo {author} {\bibfnamefont {A.~A.}\ \bibnamefont
  {El-Sharkawy}}, \bibinfo {author} {\bibfnamefont {A.~M.}\ \bibnamefont {Abou
  El-Azm}}, \bibinfo {author} {\bibfnamefont {M.~I.}\ \bibnamefont {Kenawy}},
  \bibinfo {author} {\bibfnamefont {A.~S.}\ \bibnamefont {Hillal}}, \ and\
  \bibinfo {author} {\bibfnamefont {H.~M.}\ \bibnamefont {Abu-Basha}},\
  }\bibfield  {title} {\enquote {\bibinfo {title} {{Thermophysical properties
  of polycrystalline PbS, PbSe, and PbTe in the temperature range 300--700
  K}},}\ }\href {\doibase 10.1007/BF00502357} {\bibfield  {journal} {\bibinfo
  {journal} {International Journal of Thermophysics}\ }\textbf {\bibinfo
  {volume} {4}},\ \bibinfo {pages} {261--269} (\bibinfo {year}
  {1983})}\BibitemShut {NoStop}%
\bibitem [{\citenamefont {Fan}\ \emph {et~al.}(2019)\citenamefont {Fan},
  \citenamefont {Dong}, \citenamefont {Harju},\ and\ \citenamefont
  {Ala-Nissila}}]{Fan2019prb}%
  \BibitemOpen
  \bibfield  {author} {\bibinfo {author} {\bibfnamefont {Zheyong}\ \bibnamefont
  {Fan}}, \bibinfo {author} {\bibfnamefont {Haikuan}\ \bibnamefont {Dong}},
  \bibinfo {author} {\bibfnamefont {Ari}\ \bibnamefont {Harju}}, \ and\
  \bibinfo {author} {\bibfnamefont {Tapio}\ \bibnamefont {Ala-Nissila}},\
  }\bibfield  {title} {\enquote {\bibinfo {title} {Homogeneous nonequilibrium
  molecular dynamics method for heat transport and spectral decomposition with
  many-body potentials},}\ }\href {\doibase 10.1103/PhysRevB.99.064308}
  {\bibfield  {journal} {\bibinfo  {journal} {Phys. Rev. B}\ }\textbf {\bibinfo
  {volume} {99}},\ \bibinfo {pages} {064308} (\bibinfo {year}
  {2019})}\BibitemShut {NoStop}%
\bibitem [{\citenamefont {Jiang}\ \emph {et~al.}(2021)\citenamefont {Jiang},
  \citenamefont {Zhang}, \citenamefont {Zhang},\ and\ \citenamefont
  {Wang}}]{jiang2021cpb}%
  \BibitemOpen
  \bibfield  {author} {\bibinfo {author} {\bibfnamefont {Wanrun}\ \bibnamefont
  {Jiang}}, \bibinfo {author} {\bibfnamefont {Yuzhi}\ \bibnamefont {Zhang}},
  \bibinfo {author} {\bibfnamefont {Linfeng}\ \bibnamefont {Zhang}}, \ and\
  \bibinfo {author} {\bibfnamefont {Han}\ \bibnamefont {Wang}},\ }\bibfield
  {title} {\enquote {\bibinfo {title} {{Accurate Deep Potential model for the
  Al-Cu-Mg alloy in the full concentration space}},}\ }\href {\doibase
  10.1088/1674-1056/abf134} {\bibfield  {journal} {\bibinfo  {journal} {Chinese
  Physics B}\ }\textbf {\bibinfo {volume} {30}},\ \bibinfo {pages} {050706}
  (\bibinfo {year} {2021})}\BibitemShut {NoStop}%
\bibitem [{\citenamefont {Lu}\ \emph {et~al.}(2021)\citenamefont {Lu},
  \citenamefont {Jiang}, \citenamefont {Chen}, \citenamefont {Zhang},
  \citenamefont {Jia}, \citenamefont {Wang},\ and\ \citenamefont
  {Chen}}]{lu2021dp}%
  \BibitemOpen
  \bibfield  {author} {\bibinfo {author} {\bibfnamefont {Denghui}\ \bibnamefont
  {Lu}}, \bibinfo {author} {\bibfnamefont {Wanrun}\ \bibnamefont {Jiang}},
  \bibinfo {author} {\bibfnamefont {Yixiao}\ \bibnamefont {Chen}}, \bibinfo
  {author} {\bibfnamefont {Linfeng}\ \bibnamefont {Zhang}}, \bibinfo {author}
  {\bibfnamefont {Weile}\ \bibnamefont {Jia}}, \bibinfo {author} {\bibfnamefont
  {Han}\ \bibnamefont {Wang}}, \ and\ \bibinfo {author} {\bibfnamefont {Mohan}\
  \bibnamefont {Chen}},\ }\href@noop {} {\enquote {\bibinfo {title} {{DP Train,
  then DP Compress: Model Compression in Deep Potential Molecular Dynamics}},}\
  } (\bibinfo {year} {2021}),\ \Eprint {http://arxiv.org/abs/2107.02103}
  {arXiv:2107.02103 [physics.comp-ph]} \BibitemShut {NoStop}%
\end{thebibliography}
\end{document}